\documentclass[11pt,a4paper]{article}

\usepackage{amsmath,graphicx,dsfont,enumerate,enumitem}
\usepackage{amsmath}
\usepackage{amssymb}
\usepackage{amsthm}
\usepackage{mathrsfs}
\usepackage{subfigure}
\usepackage{geometry}
\usepackage{color,xcolor}
  \definecolor{gr}{RGB}   {18,   154,   47 } 
  \definecolor{cya}{RGB}   {0,   174,   240 }
  \definecolor{gr1}{RGB}   {168,   208,   104 }
  \definecolor{orange}{RGB} { 255, 127, 0 }
  \definecolor{bl}{rgb}   {0.,   0.,   1. } 
  \definecolor{mg}{rgb}   {0.5,  0.,    0.7}
  \definecolor{yl}{RGB}   {218, 165, 32}

\usepackage{hyperref}
  \hypersetup{
      unicode=false,          
      pdftoolbar=true,        
      pdfmenubar=true,        
      pdffitwindow=false,     
      pdfstartview={FitH},    
      pdfauthor={J Garnier},     
      pdfcreator={J Garnier},   
      pdfpagemode = UseNone,   
      colorlinks=true,       
      linkcolor=red,          
      citecolor=orange,        
      filecolor=mg,      
      urlcolor=cyan,           
  }
\usepackage{siunitx}
 \usepackage{setspace,natbib}
\usepackage{setspace}

\def\R{\mathbb R}

\def\tup{\tilde{\upsilon}}

\def\sups{\upsilon^*}
\def\epsilon{\varepsilon}
\def\ds{\displaystyle}

\def\epsilon{\varepsilon}

\newtheorem{res}{Result}

\newcommand{\Div}{\mathrm{Div}}
\newcommand{\apprx}{\approx}

\usepackage{epstopdf}

\begin{document}


\title{Expansion under climate change: the genetic consequences}
\author{
Jimmy Garnier$^{\hbox{ \small{a,b,c}}}$, Mark Lewis$^{\hbox{ \small{c,d}}}$ \\
\footnotesize{$^{\hbox{a }}$Universit\'e Savoie Mont-Blanc, LAMA, F-73000 Chamb\'ery, France}\\
\footnotesize{$^{\hbox{b }}$CNRS, LAMA, F-73000 Chamb\'ery, France}\\
\footnotesize{$^{\hbox{c }}$Centre for Mathematical Biology, Department of Mathematical and Statistical Sciences,} \\
\footnotesize{University of Alberta, Edmonton, AB, Canada}\\
\footnotesize{$^{\hbox{d }}$Department of Biological Sciences, University of Alberta, AB, Canada}
}

\date{}

\maketitle


\begin{abstract}
{\small
Range expansion and range shifts are crucial population responses to climate change. Genetic consequences are not well understood but are clearly coupled to ecological dynamics that, in turn, are driven by shifting climate conditions. We model a population with a deterministic reaction--diffusion model coupled to a heterogeneous environment that develops in time due to climate change.  We decompose the resulting travelling wave solution into neutral genetic components to analyse the spatio-temporal dynamics of its genetic structure.  Our analysis shows that range expansions and range shifts under slow climate change preserve genetic diversity. This is because slow climate change creates range boundaries that promote spatial mixing of genetic components. Mathematically, the mixing leads to so-called {\em pushed} travelling wave solutions. This mixing phenomenon is not seen in spatially homogeneous environments, where range expansion reduces genetic diversity through gene surfing arising from {\em pulled} 
travelling wave solutions.  However, the preservation of diversity is diminished when climate change occurs too quickly. Using diversity indices, we show that fast expansions and range shifts erode genetic diversity more than slow range expansions and range shifts. Our study provides analytical insight into the dynamics of travelling wave solutions in heterogeneous environments. 
}

\end{abstract}

\section{Introduction}
Climate change is known to greatly modify the spatial--distribution of species at large spatial scale~\citep[see e.g. ][]{Par06}. After the last glaciers retreated, for instance, many species started to expand their range into newly emerging suitable habitats~\citep{Hew00,Plu11}. More recently, some species have started moving their range in response to their climatic niches shifting as a result of global warming~\citep{Par06,BatSta05,BreSti13}.
These processes of range shifts and range expansion in response to climate change and, in particular, mechanisms underpinning the dynamics have been analysed from both empirical to theoretical perspectives~\citep{Tra03,McIDytTra07}.
However, less is known about the spatial genetic consequences of range shifts induced by climate change~\citep{AreRay12,DaiXiaLuFu14,NulHal13,McITurWon09}.

Range shifts are frequent during climate change~\citep{Plu11,BreSti13,SamBarLin12,RouZha10}. They arise from colonisation of newly emerging suitable habitat that coincides with extirpation from areas that have become unsuitable~\citep{Par06,Par96,RooPriHal03,WalPos02}.
Colonisation at the leading edge of the range is well known to modify the patterns of neutral genetic diversity through the mechanism of gene surfing, in which neutral variants can rise to high frequency at the wavefront and propagate at the leading edge~\citep{EdmLil04,KloCur06}. Surfing results from strong genetic drift taking place on the edge of the population wave~\citep{ExcRay08,ExcFol09} because the growth rate at low density regions at the edge of the expanding population is typically higher than the growth rate for the bulk of the population~\citep{KloCur06}. The existence of surfing events can thus lead to a reduction in the genetic diversity of the newly colonised areas~\citep{KloCur06,ExcRay08,NeiMarCha75,NevPavKon09}. 
In the context of biological invasion or colonisation without climate change,  theoretical~\citep{EdmLil04,KloCur06,HalNel08,GooCooColLew14,RoqGarHamKle12} as well as empirical~\citep{HalHer07,EstBeaSen04,WhiPerHec13} studies have shown that these expansion processes generally erode the neutral genetic diversity at the leading edge of the colonisation or invasion wave.

Range shifts or range expansions under climate change are different from colonisation or invasions into favourable environments. 
Spatial variations in climate constrains the range of many species~\citep{PetOrtBar02,PeaDaw03,SchIvePra06}, producing range 
boundaries and defining climate envelopes within which species can survive. Thus, during range shifts or range expansions under climate change, the presence of moving climate envelopes can constrain the leading edge of the population in its expansion relative to the core of the population and may thus reduce its spreading speed. Under climate change, the population's spreading speed will depend on the climate velocity as well as   phenotypic characteristics of the population, such as dispersal ability and intrinsic growth rate, which define its \emph{potential spreading speed}---its spreading speed of colonisation if there were no climate constraint. The presence of climatic envelope is also known to modify the genetic drift on the edge of the species range~\citep{HilHugDyt06}. Using a stochastic simulation model, \citet{NulHal13} and \citet{DaiXiaLuFu14} have numerically analysed the surfing phenomenon in presence of climate constraints. More precisely, using a coalescent model with 
a backward-time 
approach, they analysed the distribution of the time to common ancestry for individuals sampled along the expansion wave to derive the position of successful surfers in the wave. Significant differences in the diversity patterns were found between populations that experience  range expansion under climate change and populations that expand their range freely. Combining their stochastic approach with the framework of reaction-diffusion equations, they were able to connect their numerical findings to analytical formulae. 
Among other things, they concluded that surfing is possible in deterministic reaction-diffusion equations with shifting heterogeneous media. Our goal is to investigate how the climate constraint determines genetic diversity in expanding or shifting range and to do so in a context that is broader than  the surfing phenomenon.

During a range shift, there is also range retraction at the rear edge to consider. This process can threaten both the survival of the species~\citep{BerDie09} and its neutral genetic diversity~\citep{AreRay12}. For instance, rapid climate change can cause extinction if the species cannot track the moving climate envelope because of limited dispersal~\citep{SchVerVos11}.

This paper investigates how the interplay between a population, the climate velocity, and the climate envelope size determines neutral genetic diversity patterns during a climate change. It hinges on the question as to how  genetic fractions evolve inside a one--dimensional wave travelling in response to climate change. Clearly, the fraction at the leading edge of the wave should have a higher probability of surviving and surfing on the wave~\citep{McITurWon09}. On the other hand, this fraction may not survive if it travels too far forward into inhospitable habitat. This would enhance the predominance of remaining genetic fractions from the bulk of the wave as they emerge at the leading edge of the wave. We show that this loss into inhospitable habitat promotes genetic diversity at the leading edge of the wave. The climate velocity also plays an important role in the balance between population spread and loss. In the range shift scenario, extirpation at the rear edge of the wave can also play a role causing 
loss of genetic fractions~\citep{AreRay12}, thus eroding genetic diversity.

\begin{figure}
    \centering
\includegraphics[width=129mm]{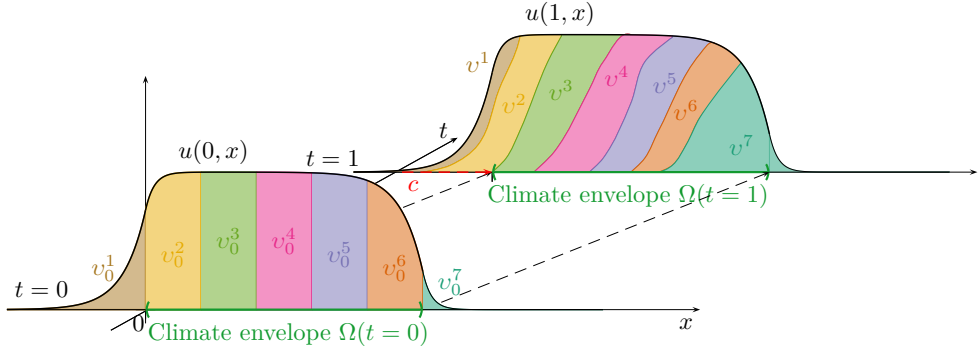}
\caption{A schematic representation of a wave $u(t,x)$ described by~\eqref{eq:RDFS} and composed of seven fractions, travelling to the right to track the climate envelope $\Omega(t),$ shifting with a constant speed $c$ in a one-dimensional environment. Each neutral fraction is depicted with a different colour and with a thickness corresponding, at each position $x$, to the density $\upsilon^i$ of the fraction.}\label{fig:environment}
\end{figure}

Following the framework provided in~\citep{BerDie09,BerRos08,PotLew04}, we focus on the deterministic  one-dimensional heterogeneous reaction--diffusion equations with forced speed of the form
\begin{equation}\label{eq:RDFS}
    \partial_t u(t,x) = D\, \partial_{xx} u(t,x) +  f(x-ct,u(t,x)),  \quad t>0, \ x\in(-\infty,\infty),
\end{equation}
where $u=u(t,x)$ represents the population density (of genes or haploid individuals) at time $t$ and location $x$. This density changes in time under the joint effects of local dispersal, accounted for by the diffusion term with diffusivity $D>0$, and local reproduction, described by the growth function $f$. Since the climate conditions are not uniform in space, some regions are more favourable than others for the species. The growth function is therefore heterogeneous in space, and the term $f(x - ct,u(t,x))$ expresses the suitability profile of this environment. In the absence of climate change, we assume that this environment is composed of a bounded climate envelope of a suitable habitat, surrounded by unfavourable regions where the population tends to go extinct. Moreover, we only assume negative density dependence in the climate envelope, i.e., we do not take any Allee effect into account. The growth function is essentially logistic and the precise assumption on the growth term $f$ is described in the 
next section. We reflect climate change in our model by assuming that the climate envelope moves rightward in space at a constant speed $c>0$~(see Fig.~\ref{fig:environment}). 

Since Skellam's work~\citep{Ske51}, these models have commonly been used to explore population range expansions. On the other hand, \citet{Nag75,Nag80,Nag00} have also used these models to describe the change of allele frequencies in population with varying density and dispersal.  The mathematical modelling novelty of this current paper, comes from combining these two approaches to describe the evolution of the inside structure of wave-like solutions associated with a shifting climate envelope. From a mathematical view point, our analysis extend the early work of~\citet{RoqGarHamKle12} to more realistic landscape which evolves in space and time. 

The solutions $u$ to the initial value problem for~\eqref{eq:RDFS} converge to a travelling wave solution under reasonable assumptions (to be given later) on the function $f$, the speed $c$, the diffusivity $D$, and the initial population density $u_0$~\citep{BerRos08}. In the context of climate change, these solutions describe a population that shifts its range at the same speed $c$ as the climate envelope and with a constant density profile $U_c$. Thus, the population density is of the form $u(t,x)=U_c(x - ct)$.

Following the pioneering works of~\citet{Nag80,Nag00}, we assume that the population wave is initially composed of several spatially distinct neutral fractions. We provide a mathematical analysis of the spatio-temporal dynamics of these fractions, i.e., we analyse how the density of  each fraction develops over time at each position in the travelling wave.  This analysis contrasts with classical approaches on reaction--diffuion with heterogeneous environment which mainly focus on the dynamics of the total waves rather than the dynamics of the neutral fractions~\citep[see][for a review]{Xin00}. In this paper, we provide mathematical insights into the following theoretical issues: 
\begin{description}
    \item[(Q1)]~{How do the densities of the various fractions evolve in a travelling wave generated by a reaction--diffusion model with an environment that changes with time and space? {Are these travelling waves solutions of heterogeneous reaction--diffusion pulled or pushed?};}
    \item[(Q2)]~{How does the presence of climate change modify the fraction densities inside a travelling wave?
    Does it enhance or reduce the genetic diversity in a colonisation front?}
    \item[(Q3)]~{How do the species' spreading potential, the climate velocity, and the climate envelope size modify the genetic patterns inside a range shifting in response to climate change?}
\end{description}

\section{The model, main assumptions, and classical results}
We assume that the population $u$ is composed of several spatially distinct neutral fractions $\upsilon^i$ (see Fig.~\ref{fig:environment}). To analyse the dynamics of these fractions, we extend  the mathematical notion of~\emph{inside dynamics} of a solution, introduced in~\citep{RoqGarHamKle12,GarGilHamRoq12} and described in what follow. 
At time $t=0$
\begin{equation}\label{eq:v0}
u_0(x):=u(0,x)=\ds\sum_{i = 1}^I {\upsilon_0^i(x)} \quad \mbox{with} \ \upsilon_0^i\geq0 ,
\end{equation}
where $I$ is the number of fractions.
We assume that the genes (or individuals) in each fraction differ only by their positions and their alleles (or their labels), while their dispersal and growth capabilities are the same as those for the entire population $u$. More precisely, the fraction density $\upsilon^i(t,x)$ grows according to $f(x - ct,u(t,x))$ rescaled by the fraction frequency or allele frequency $\upsilon^i(t,x)/u(t,x)$. Each density $\upsilon^i$ therefore satisfies the following equations:
\begin{equation}\label{eq:syst^i}
   \left\{\begin{array}{l}
           \ds \partial_t \upsilon^i  =  D\partial_{xx}\upsilon^i+ \frac{f(x - ct,u)}{u} \upsilon^i, \ t>0, \ x\in(-\infty,\infty), \\[6pt]
            \upsilon^i(0,x)  =  \upsilon_0^i(x), \ x\in\R.
          \end{array}\right.
\end{equation}
As expected, it follows from the uniqueness of the solution to the initial value problem associated with~\eqref{eq:RDFS} that the sum of the fraction densities $\upsilon^i$ is equal to the population density $u$.

\subsection{Diversity measure}
Our decomposition approach gives a conceptual framework for describing and analysing the diversity dynamics of a population that is shifting its range due to climate change. More precisely, for each $i \in \{1,\ldots,I\}$ we denote the frequency of fraction  $i$ in the population $u$ by $p^i:=\upsilon^i/u$. We can define at each location $x\in\R$ and time $t>0$ a family of diversity measures $\Div^q(t,x)$ in terms of $p^i$, where $q\in[0,\infty]$; see~\citep{LeiCob12}. These indices quantify the local diversity according to the fraction frequency $p^i$ and the sensitivity parameter $q$.
The diversity of order $q$ is defined for any location $x\in\R$ and time $t>0$ by
\begin{equation}\label{eq:Divq}
\begin{array}{l}
   \ds  \Div^q=\left(\sum_{i=1}^I{(p^i)^q}\right)^{\frac{1}{1-q}} \quad \mbox{for} \ q\neq1,\infty , \\[6pt]
 \ds \Div^1=\prod_{i=1}^{I}{\frac{1}{(p^i)^{p^i}}}, \quad \mbox{and} \quad
 \Div^\infty=\frac{1}{\ds\max_{i\in\{1,\dots,I\}}{p^i}}.
\end{array}
\end{equation}
The sensitivity parameter $q$ controls the relative emphasis placed on common and rare fractions and indicates how much significance is attached to fraction frequency. For example, at one extreme $(q=0)$, the index $\Div^0$ corresponds to species richness, attaching as much significance to rare fractions as to common ones. At the other extreme $(q=\infty)$, the index of~\citet{BerPar70} depends only on the most frequent fraction, rarer fractions being ignored altogether. The exponential of the Shannon index~\citep{Sha48}, corresponding to $q = 1$, quantifies the entropy (or uncertainty) in predicting the fraction identity of a gene chosen at random from a location $x$ at time $t$. The inverse of the Simpson index~\citep{Sim49}, corresponding to $q = 2$, describes the probability that two individuals sampled randomly at location $x$ and time $t$ belong to the same fraction. For any $q$, a high index of diversity of order $q$ indicates high diversity or a true evenness in the population: $\
Div^q$ is maximal when all the fraction frequencies are equal, i.e., when $p^1=\cdots=p^I=1/I$.

\subsection{Growth function and climate envelope}
We consider a one-dimensional environment $(-\infty,\infty)$ composed of one climate envelope, surrounded by unfavourable habitat. Due to climate change, the advantageous patch moves with fixed speed $c\geq0$ to the right. In the favourable habitat, the population can reproduce, while in the unfavourable regions there is no reproduction and the population dies at a rate $d>0$. 
We focus on two scenarios. 

\paragraph*{Scenario~1: range expansion under climate change.}~First, we consider a habitat expansion due to climate change, for instance expansion due to glacier retreat~\citep{Plu11}.  In this case we are only interested in the colonisation part and we neglect the possible retraction at the rear edge of the range. Thus the climate envelope size is infinite ($L=\infty$) and the growth function is defined by
\begin{equation}\label{eq:f_infty}
 f(x,u)=\left\{\begin{array}{ll}
                \ds r\,u\Big(1-\frac{u}{K}\Big) & \hbox{if } \ x\in(-\infty,0),\\
                -d\,u & \hbox{if } \ x\in [0,\infty),
               \end{array}\right.
\end{equation}
where $r>0$ is the intrinsic growth rate of the population.  In the favourable region, we only take into account negative density dependence where $K$ describes the carrying capacity of the climate envelope. Thus the per capita growth rate $g(x,u)=f(x,u)/u$ is a decreasing function of $u$ at location $x$ inside the climate envelope. Note that no Allee effect is involved in the growth process.
Without loss of generality, we may rescale time so as to fix $d=1$ and the solution $u$ so as to fix $K=1.$

\paragraph*{Scenario~2: range shift.}~Secondly, we investigate the range shift scenario in which the climate envelope is assumed to be of finite size $L>0.$ The specific growth function $f$ takes the form
\begin{equation}\label{eq:f_shift}
 f(x,u)=\left\{\begin{array}{ll}
                \ds r\,u\Big(1-\frac{u}{K}\Big) & \hbox{if } \ x\in(-L,0),\\
                -d\,u & \hbox{if } \ x\in(-\infty,-L]\cup[0,\infty),
               \end{array}\right.
\end{equation}

\subsection{Travelling wave solutions and persistence}
In each scenario, if the climate envelope shifts at speed $c\geq0$, the travelling wave solution satisfies $u(t,x)=U_c(x-ct)$. Substituting this expression into~\eqref{eq:RDFS}, we see that the profile $U_c$ of the travelling wave solves the elliptic equation
\begin{equation}\label{eq:U_1D}
 D\,U_c'' +c\,U_c' +f(x,U_c)=0 \quad \mbox{with} \ -\infty < x  < \infty \ \mbox{and} \ U_c \ \mbox{in} \ (0,1) .
\end{equation}
These travelling waves propagate in the direction of the climate change (from left to right). In the range expansion scenario, they take the form of a decreasing wave describing the invasion into an environment where the species is not yet present with a constant speed $c$ and a constant density profile $U_c.$ In the range shift scenario they move with a shifting climate envelope and so the profile $U_c$ is bell--shaped because there is expansion at the leading edge of the climate envelope and contraction at the rear edge (see Fig.~\ref{fig:environment}). One can note that increasing contraction constraint tends to flatten the trailing edge slope of the profile (see Fig.~\ref{fig:vi_1D_RS}).

The existence of these solutions depends critically on the dispersal ability $D$ of the species and on the environmental parameters ($c$, $L$, and $r$). In the range expansion scenario,~\citet{BerDie09} and~\citet{PotLew04} showed that a unique travelling wave solution exists if and only if $c < c^*(r,D),$ where $c^*(r,D)$ is the \emph{potential spreading speed}, defined by
\begin{equation}\label{eq:c_star}
c^*(r,D):=2\sqrt{rD}.
\end{equation}
In the range shift scenario, they proved that the existence holds if and only if both $c < c^*(r,D),$ and $L > L^*(c^*,c)$, a critical envelope size, defined by
\begin{equation}\label{eq:Lstar_1D}
 L^*(c^*,c):=\ds \frac{\frac{c^*}{r}}{\sqrt{1-\big(\frac{c}{c^*}\big)^2}}
             \arctan{\left( \frac{\sqrt{\frac{d}{r}+\big(\frac{c}{c^*}\big)^2}}{\sqrt{1-\big(\frac{c}{c^*}\big)^2}} \right)}.
\end{equation}
The potential spreading speed $c^*$ defined by~\eqref{eq:c_star} is the speed at which the species would spread in absence of climate constraint. Note that it only depends on the phenotype characteristics $r$ and $D.$

\begin{figure}
\centering
\includegraphics[width=129mm]{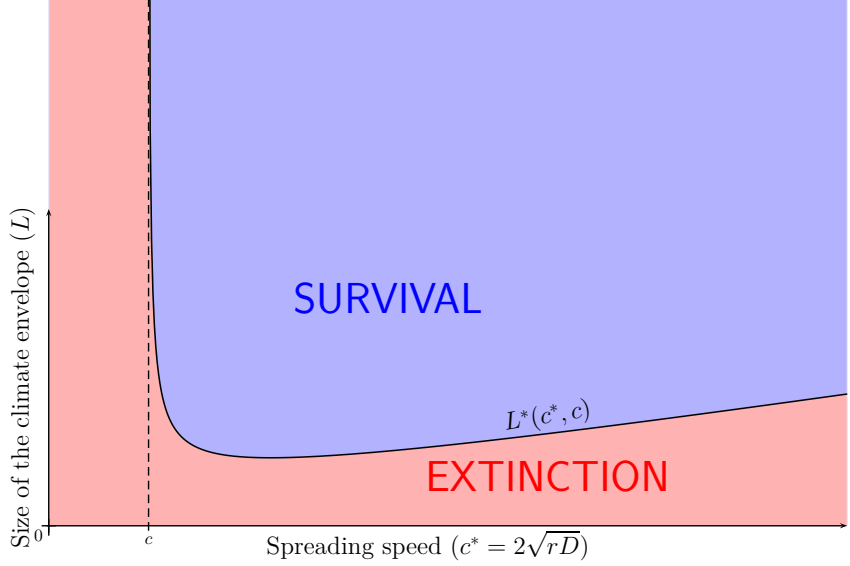}
\caption{Schematic representation of the behaviour of the population density $u$ as a function of the size of the climate envelope $L$ and the potential spreading speed $c^*=2\sqrt{rD}$. Only a population in a wide climate envelope or with a high potential spreading speed can survive climate change.}\label{fig:survival_Lstar}
\end{figure}

\citet{BerDie09} showed that, for any biologically reasonable initial density (mathematically described as a nontrivial bounded initial condition $u_0$), the population density $u$, satisfying~\eqref{eq:RDFS}, will persist under precisely the same conditions as above, i.e., if and only if $c < c^*(r,D)$ in scenario~1 and if in addition $L > L^*(c^*,c)$ in scenario~2 (Fig.~\ref{fig:survival_Lstar}). In any case, the density $u$ converges to the travelling wave solution of~\eqref{eq:RDFS}.

\section{Range shifts and range expansions under climate change preserve richness of neutral genetic fraction}
Henceforth, we assume that the entire population $u$ satisfies $u(t,x)=U_c(x-ct)$ with $U_c$ solving~\eqref{eq:U_1D}. Thus, the population has already attained its asymptotic travelling wave profile. 
We consider the population $u$ to be composed of $I$ neutral fractions, the $i$th fraction having density $\upsilon^i$ satisfying~\eqref{eq:syst^i}, and study the dynamics of each neutral fraction density $\upsilon^i(t,x)$ and of the associated diversity measures $\Div^q(t,x)$.

\subsection*{Dynamics of the neutral fractions inside the population}
For both scenarios, we consider an arbitrary fraction density $\upsilon^i$ satisfying~\eqref{eq:syst^i} and, for simplicity, refer to it as $\upsilon$ henceforth. The density of this fraction therefore satisfies
\begin{equation}\label{eq:RDFS_1D_syst^i}
\begin{array}{l}
\ds \partial_t \upsilon=  D\,\partial_{xx} \upsilon +  g( x-ct, U_c(x-ct) ) \,\upsilon ,  \quad t>0, \ x\in(-\infty,+\infty).
\end{array}
\end{equation}
Moreover, at time $t=0$, the entire population satisfies $u_0=U_c(x)$ and the fraction $\upsilon_0(x)=\upsilon(0,x)$ corresponds to a portion of the quantity $U_c(x)$. Thus, $0\leq \upsilon(0,x)\leq U_c(x)$ for all $x\in\R$. From a biological perspective, this means that we consider the spatio-temporal dynamics of the neutral genetic diversity in an ongoing range shift or range expansion due to climate change.

Using properties of the population dispersal ability $D$, the speed $c$ of the climate envelope, and the characteristics of the profile $U_c$, we are able to describe the dynamics of the fraction density $\upsilon$ in the whole wave of propagation. 

\begin{res}\label{theo:pushed}
For any initial condition $\upsilon_0$, the density $\upsilon$ of the fraction converges uniformly in space (as $t\to\infty$) to a proportion $p(\upsilon_0)$ of the total population $u(t,x)$, that is, $\upsilon(t,x+ct)\to p(\upsilon_0)U_c(x)$ uniformly on compact sets as $t\to\infty$. The proportion $p(\upsilon_0)$ can be computed explicitly as follows:
\begin{equation}\label{eq:p}
  p(\upsilon_0)=\ds\frac{\ds\int_{-\infty}^{\infty}{\hspace{-2mm}\upsilon_0(x)\,U_c(x)\,e^{\frac{c}{D}x}\,dx}}{\ds\int_{-\infty}^{\infty}{\hspace{-2mm} U_c^2(x)\,e^{\frac{c}{D}x}\,dx}}.
\end{equation}   
\end{res}

Result~\ref{theo:pushed}, proved in section~\ref{sec:proof}, { describes the large time dynamics of the fractions inside the travelling wave generated by a reaction--diffusion model with an environment that changes with time and space. More precisely,} it shows that in each scenario, any fraction initially represented in the population $u$ (i.e., having initial density $\upsilon_0 > 0$) can follow the population and the climate envelope. 

The formula~\eqref{eq:p} has practical significance because it provides a precise information regarding the contribution of the ancestral population and the origin of the individuals that compose the wave at large time. Indeed, we have shown that, in the moving frame having speed $c$, the profile $\upsilon(t,x+ct)$ of any fraction tends to resemble the profile $U_c$ of the entire population with a scaling factor $p(\upsilon_0)$ that depends only on the initial density $\upsilon_0$. This factor $p(\upsilon_0)$ represents the contribution of the fraction $\upsilon$ at large time which is by definition the reproductive value of this fraction~\citep{BarEth11}.

Our result~\ref{theo:pushed} also shows that any fraction is represented at the leading edge and contribute to push the wave forward. 
\citet{RoqGarHamKle12} and~\citet{GarGilHamRoq12} have shown that this inside dynamics characterised the pushed wave, a notion introduced by~\citet{Sto76}. The travelling wave solution of the heterogeneous reaction--diffusion model~\eqref{eq:RDFS} is thus \emph{pushed} in this sense. { From a mathematical perspective, this result generalises the notion of pulled/pushed waves to a broader class of equations. In addition, it shows how the presence of spatial heterogeneity not only modifies the spreading speed of travelling wave solution of reaction--diffusion model but changes the dynamics within the wave itself.}

The pushed nature of this travelling wave sharply contrasts with the pulled behaviour exhibited by inside dynamics of travelling wave in a favourable homogeneous environment. For example, if we extend the good habitat to the entire environment, the growth function $f$ is then defined by $f(x,u):=r\,u(1-u)$ for all $x\in(-\infty,\infty)$. In this case, travelling wave solutions of the homogeneous equation
\begin{equation}\label{eq:RD}
\ds  \partial_t u= D \partial_{xx}u  + r\,u(1-u), \ t>0 \hbox{ and } x\in(-\infty,\infty)
\end{equation} 
are \emph{pulled} in the sense that only the fraction initially well represented at the leading edge of the travelling wave are present in the wave of expansion as $t\to+\infty$~\citep{HalNel08,GarGilHamRoq12,Sto76}. From an ecological and genetic perspective, this points out a major difference between expansion under climate change and expansion without climate constraint. Thus, we can expect that the genetic diversity patterns induced by an expansion caused by climate change will be very different than patterns induced by biological invasion without any climate constraint. Moreover, we see that genetic diversity can be preserved in a range expansion in response to climate change~\citep[see also ][]{Plu11,DaiXiaLuFu14,NulHal13}.

It is instructive to consider a fraction starting with an initial position $\alpha$ on the wave, $\upsilon^\alpha_0=U_c\cdot \delta_\alpha$, for some $\alpha\in\R$ and $\delta_\alpha$ the Dirac mass at location $\alpha.$ The reproductive value of this fraction, $p(\alpha):=p(\upsilon^\alpha_0)$,  which describes the relative contribution to the wave of the individuals with an initial position $\alpha$ on the wave, can be computed by the formula
\begin{equation}\label{eq:p_alpha}
   p(\alpha) = \ds\frac{\ds U^2_c(\alpha)\,e^{\frac{c}{D}\alpha}}{\ds\int_{-\infty}^{\infty}{\hspace{-2mm} U_c^2(x)\,e^{\frac{c}{D}x}\,dx}}.
\end{equation}
{ In the context of an expanding population facing an Allee effect, a similar formula, 
replacing $U_c$ with the solution of stochastic model, has provided a good fit for the probability of gene surfing in these models~\citep{HalNel08,BarEthKelVeb13,DurWai16}. The presence of an Allee effect is known to enhance genetic diversity along colonisation waves~\citep{RoqGarHamKle12}. Our results reveals an influence of the climate envelope that is analogous to that of an Allee effect on the genetic diversity inside a colonisation wave. This analogy will be discussed in section~\ref{sec:discusion}. }


{
In our model,  the travelling profile $U_c$ only depends on the carrying capacity $K$ by a scaling factor $K.$ Thus we can deduce from  formula~\eqref{eq:p_alpha} that the carrying capacity has no influence on the relative contribution of the fractions. This result sheds a new light on the limited influence of carrying capacity on gene fractions. Indeed, by way of contrast~\citet{NulHal13} have shown  that decreased carrying capacity reduces genetic diversity along the colonisation front because it decreases the population density clines at the leading edge of the travelling waves. 
The formula~\eqref{eq:p_alpha} enhances the idea that the relative contributions of the fractions and thus the genetic diversity of the travelling wave depend only on the population density clines at climate envelope frontier, that is the steepness of the travelling wave profile $U_c.$
}

From an ecological perspective, our result shows that genetic richness, corresponding to the number of genetic fractions with positive density at location $x$ and time $t$, can be preserved during climate change. For instance, if the population $u$ is composed of $I$ fractions whose densities $\upsilon^i$ satisfy~\eqref{eq:RDFS_1D_syst^i} and $u(0,x)=\sum_{i=1}^{I}{\upsilon^i_0}$ for all $x\in\R$, Result~\ref{theo:pushed} and the maximum principle imply that the densities $\upsilon^i$ remain positive everywhere for all time $t>0$. Therefore, the genetic richness remains uniformly equal to $I$, the number of fractions initially present inside the population. However, the evenness of the gene population is modified by climate change, since over time the densities $\upsilon^i$ approach an asymptotic value of $p(\upsilon_0^i)U_c$, where $p(\upsilon^i_0)$ is defined by~\eqref{eq:p}. Moreover, the proportions $p(\upsilon_0^i)$ depend crucially on the speed $c$ of the climate shift and the profile $U_c$ of the 
travelling wave. Therefore, both climate change and the response of the entire population affect evenness of the genetic diversity.

\section{Rapid expansion always erodes diversity}
We initially focus on the expansion phenomena which corresponds to scenario~1 and we compare the effects of range expansion on genetic diversity with and without climatic change. We consider a biological invasion spreading into a homogeneous environment without climate change as a baseline case. Here, the population density is represented by the model~\eqref{eq:RD} and it spreads at its potential spreading speed, $c^*:=2\sqrt{r D}$. Secondly, we focus on range expansion under climate change (scenario~1) where the rate of habitat expansion is $c>c^*.$ So as to be able to compare outcomes, we assume the population's diffusivity $D$ and growth rate $r$ in the climate envelope to be fixed the same as the invasion case.
Our investigation employs numerical computations, as described in the next section. This allow us to explore whether the analytical Result~\ref{theo:pushed} continues to hold even when the  initial condition $u_0(x)=U_c(x)$ does not correspond to an already established travelling wave.

\subsection*{Numerical computations}
We explore numerically whether the Result~\ref{theo:pushed} regarding the asymptotic proportions for each fraction remains qualitatively true even when $u_0$ is compactly supported step function (see Fig.~\ref{fig:vi_1D_init}), before the solution $u$ has attained a travelling profile.
We consider a population consisting of $I=7$ fractions with densities $\upsilon^i$ ($i \in \{1,\ldots,7\}$). At time $t=0$, the whole population satisfies $u_0(x):=\mathds{1}_{(-\infty,2)}$, where $\mathds{1}_{(-\infty,2)}$ is the characteristic function of the interval $(-\infty,2)$, and the fraction densities satisfy $\upsilon^1_0=\mathds{1}_{(-\infty,-10]}$, $\upsilon^i_0=\mathds{1}_{(x_{i-1},x_{i}]}$ for $i=2,\dots,I-1$, and $\upsilon^I_0=\mathds{1}_{(0,2)}$, where the $x_i$ are evenly spaced points with $-10 = x_1 < x_2 < \cdots < x_{I-1} =0$. We numerically solve~\eqref{eq:RDFS} with the climate velocities $c=2$ and $c=5$, and we compute the solution to~\eqref{eq:RD}. Each fraction density $\upsilon^i$ corresponds to a different colour. For each fraction, the diffusivity is equal to $D=2$ and the per capita growth rate is equal to $r=10$, so the  potential spreading speed is $c^*=2\sqrt{rD}\apprx 6.32$. In any case, we observed that the solution $u(t,x)$ rapidly converges to a travelling wave profile (
dashed curves in Fig.~\ref{fig:vi_1D}).

\

\subsubsection*{Invasion without climate change.} 
Fig.~\ref{fig:vi_1D_UE} shows the evolution of the spatial structure of the solution $u(t,x)$ to~\eqref{eq:RD} without climate change.
As described in~\citep{RoqGarHamKle12}, only the fraction at the leading edge of the front follows the population expansion to the right. This indicates that range expansion leads to a strong erosion of diversity, caused by the demographic advantage of isolated individuals ahead of the colonisation front. In addition, we observe that genetic diversity has a vertical spatial pattern that indicates that the population is highly structured.

\

\subsubsection*{Range expansion under climate change.} 
\begin{figure}
\centering
\subfigure[Initial condition]{\includegraphics[width=5cm]{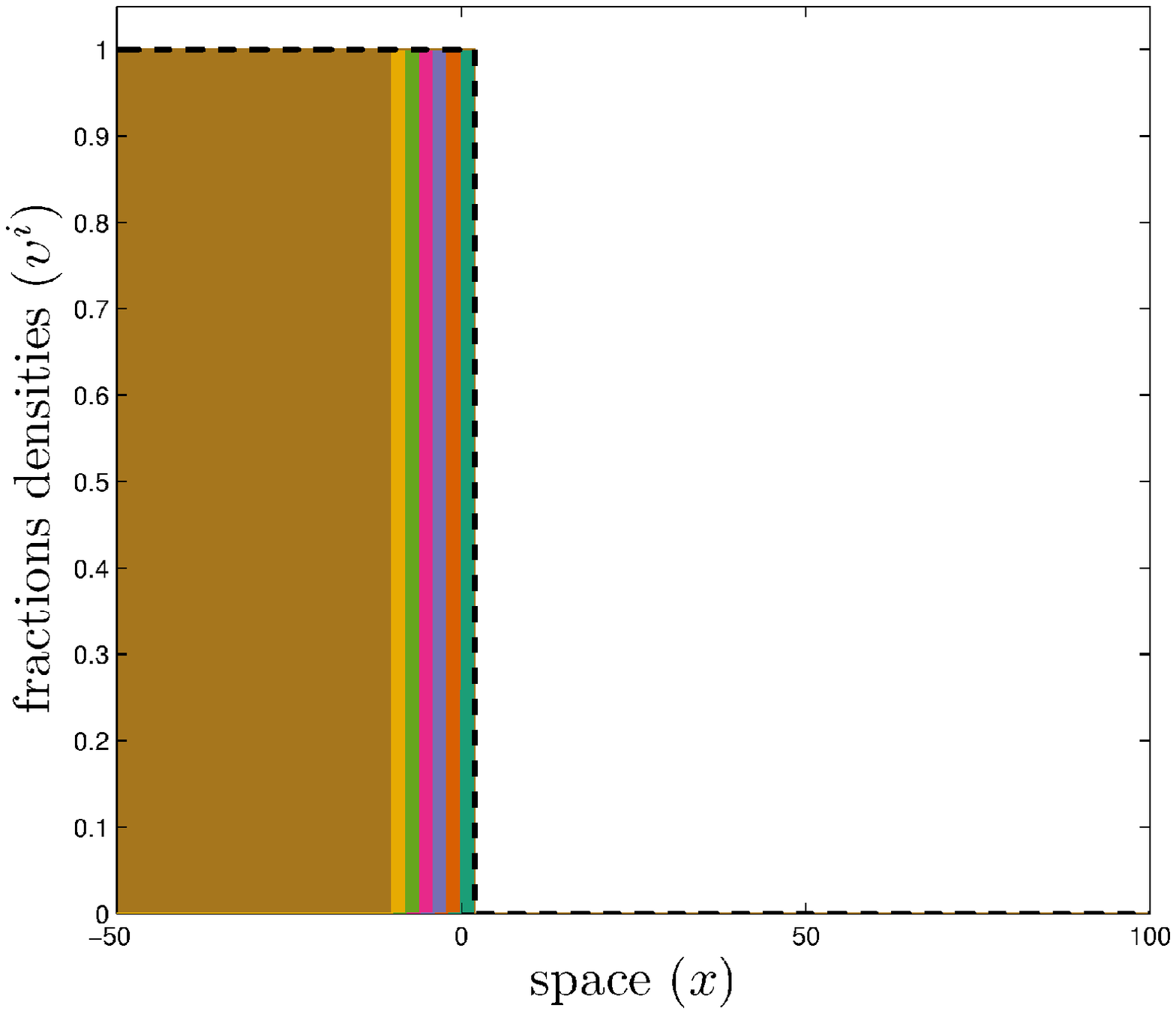} \label{fig:vi_1D_init}}\qquad
\subfigure[Invasion without climate change $(c^*\apprx 8.94)$]{\includegraphics[width=5cm]{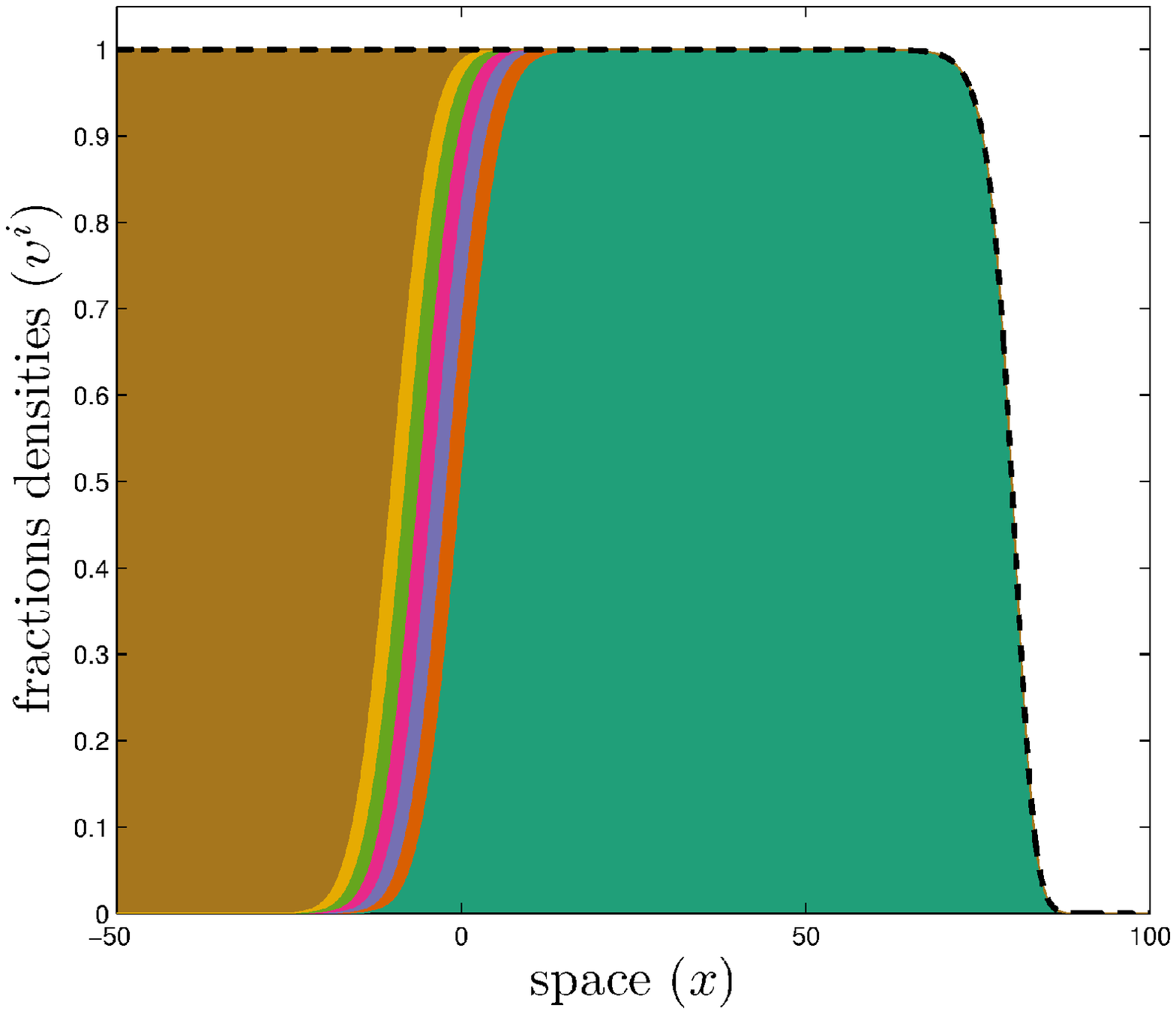}\label{fig:vi_1D_UE}} \\
\subfigure[Range expansion under slow climate change $(c=2)$]{\includegraphics[width=5cm]{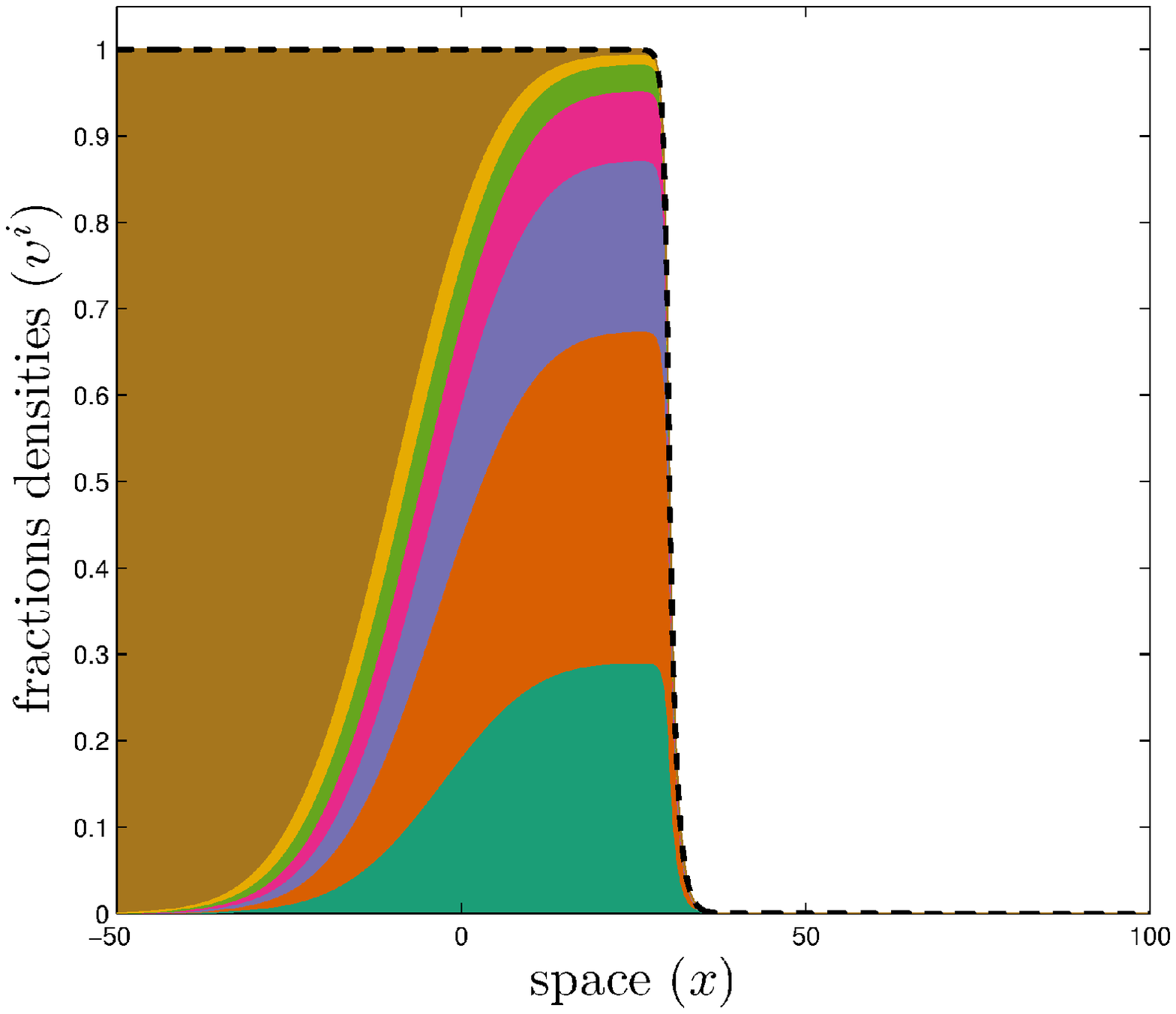}\label{fig:vi_1D_RE_slow}}\qquad
\subfigure[Range expansion under fast climate change $(c=5)$]{\includegraphics[width=5cm]{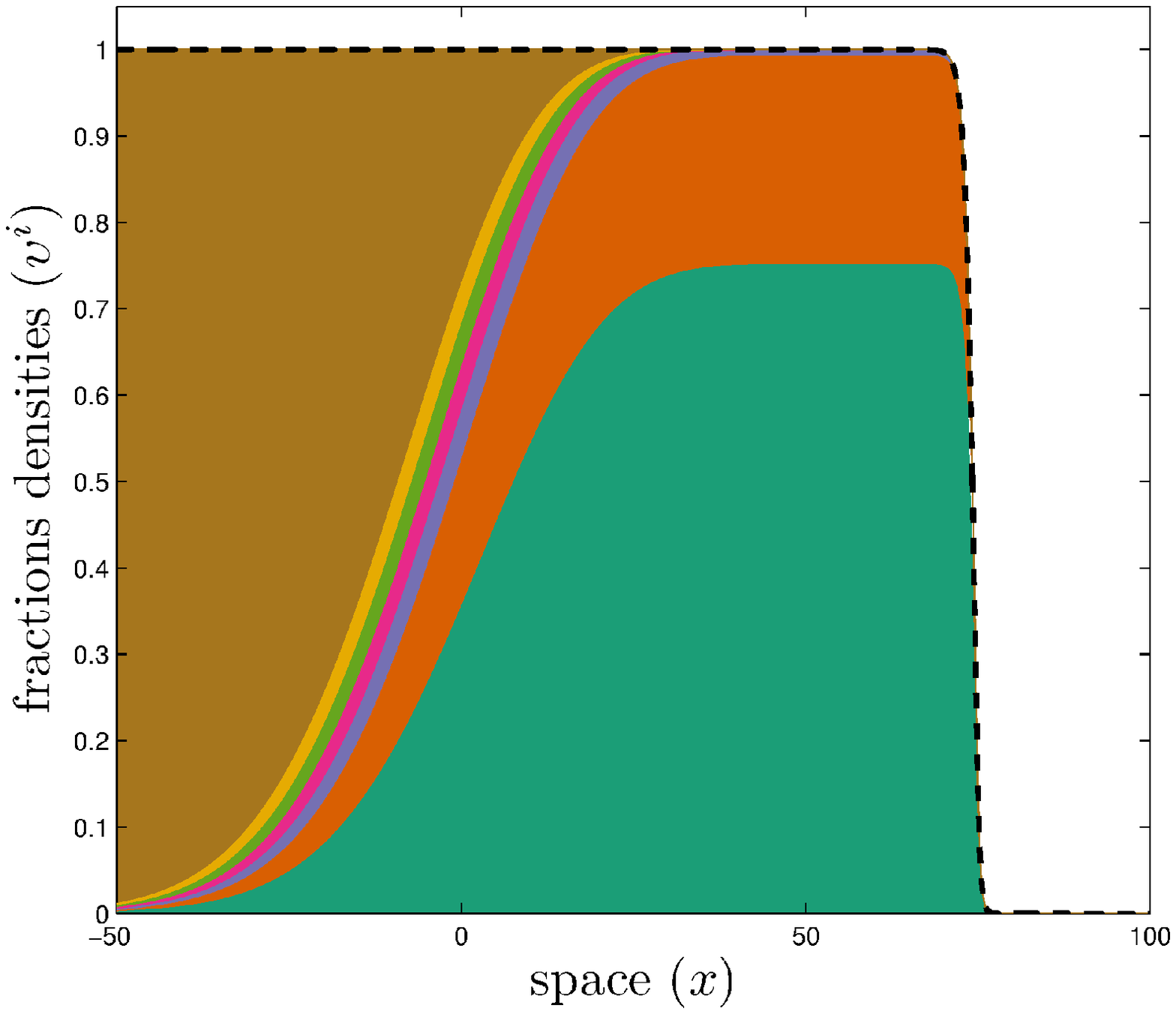}\label{fig:vi_1D_RE_fast}}
\caption{ The progress of the spatial structure of the solutions $u(t,x)$ to~\eqref{eq:RD} and~\eqref{eq:RDFS} with~\eqref{eq:f_infty}. (a) Initial structure of the population, (b) inside dynamics of the solution $u$ to~\eqref{eq:RD} spreading at speed $c^*=2\sqrt{r\,D}\apprx 8.94$ in a favourable homogeneous environment, (c)--(d) inside dynamics of the solution $u$ to~\eqref{eq:RDFS} in a moving environment under range expansion (scenario~1). In each case, the dashed black curve corresponds to the profile $U_c$ of the stable travelling wave solution of~\eqref{eq:U_1D} with~\eqref{eq:f_infty}. Climate constraints maintain diversity richness during expansion but increasing climate velocity modifies diversity evenness.
 }\label{fig:vi_1D}
\end{figure}

In the presence of range expansion under climate change (scenario~1), the situation is very different. We observe from Fig.~\ref{fig:vi_1D}(c),(d) that range boundary induced by climate change promotes diversity within at the leading edge of the travelling wave of colonisation. Even if the rightmost fraction remains well represented, any fraction initially presented inside the bulk of the population is conserved at the leading edge of the colonisation wave.

However, the faster the climate change is, the less fractions situated deep in the core of the population contribute to the wave. 
This indicates that fast range expansion due to fast climate change or biological invasion leads to a more uneven distribution of the fractions than slow range expansion in response to slow climate change. 
 
\section{Loss of diversity due to fast climate change can be offset by a high potential spreading speed}
Our previous results have shown that an increase in the climate velocity tends to reduce genetic diversity. 
We now investigate the range shift scenario~2, which combines colonisation of newly available habitat 
and extirpation at newly unsuitable habitat. We show that the intensity of erosion induced by fast climate change depends crucially on the ratio between the size of the climate envelope and the potential spreading speed of the species. 
Since these factors play an important role in the survival of the population (see Fig.~\ref{fig:survival_Lstar}) as well as the shape of the travelling wave $U_c$, they should also modify the genetic fraction and the diversity profiles $\Div^q.$

\subsection*{Numerical computations}
\begin{figure}
\centering
\includegraphics[width=130mm]{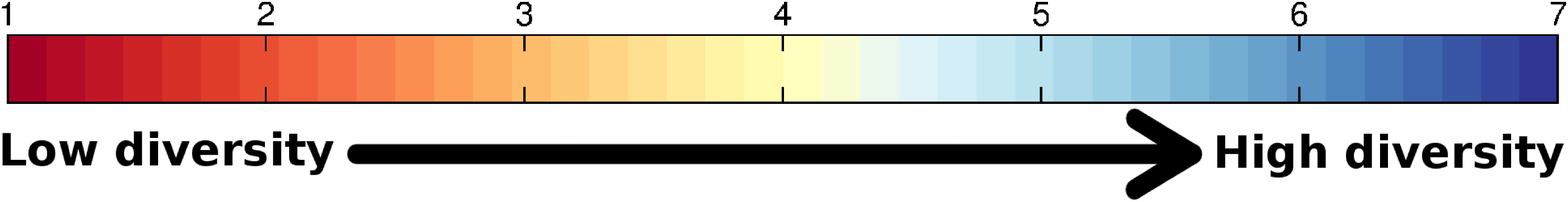}\\
\subfigure[Fixed climate envelope $(c=0)$]{
\includegraphics[width=5cm]{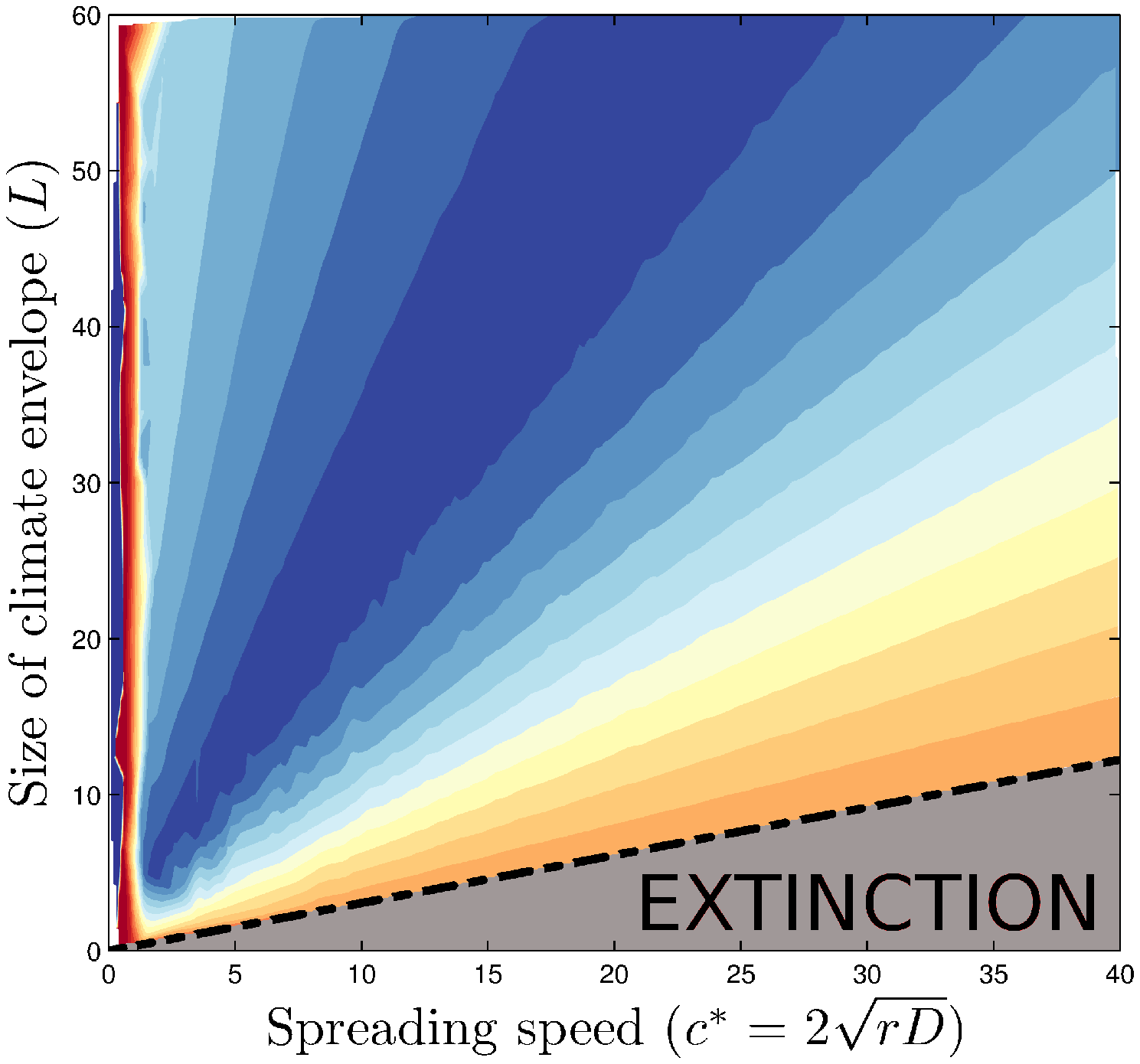}}
\subfigure[Slow climate change $(c=2)$]{
\includegraphics[width=5cm]{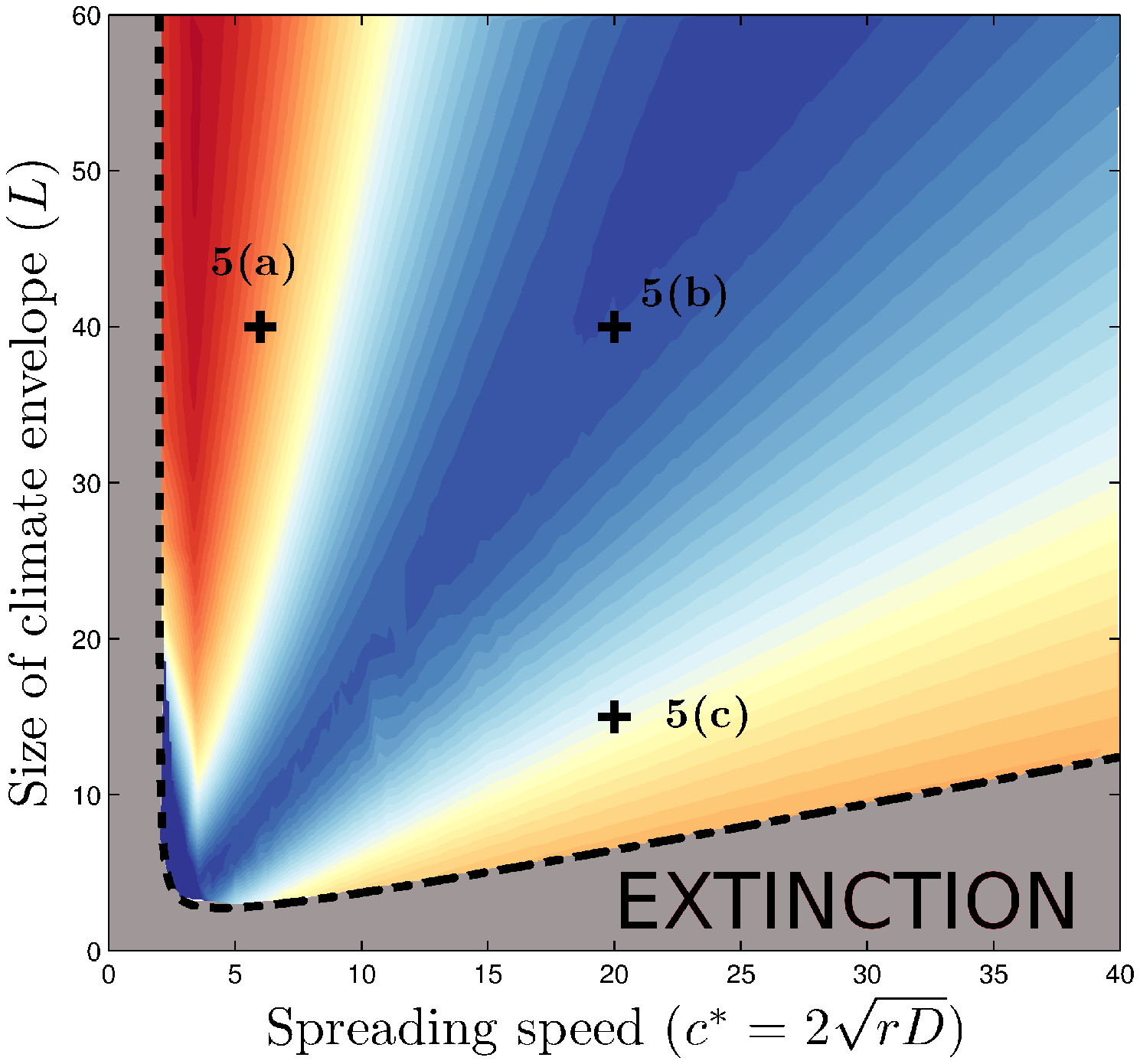}}
\subfigure[Fast climate change $(c=5)$]{
\includegraphics[width=5cm]{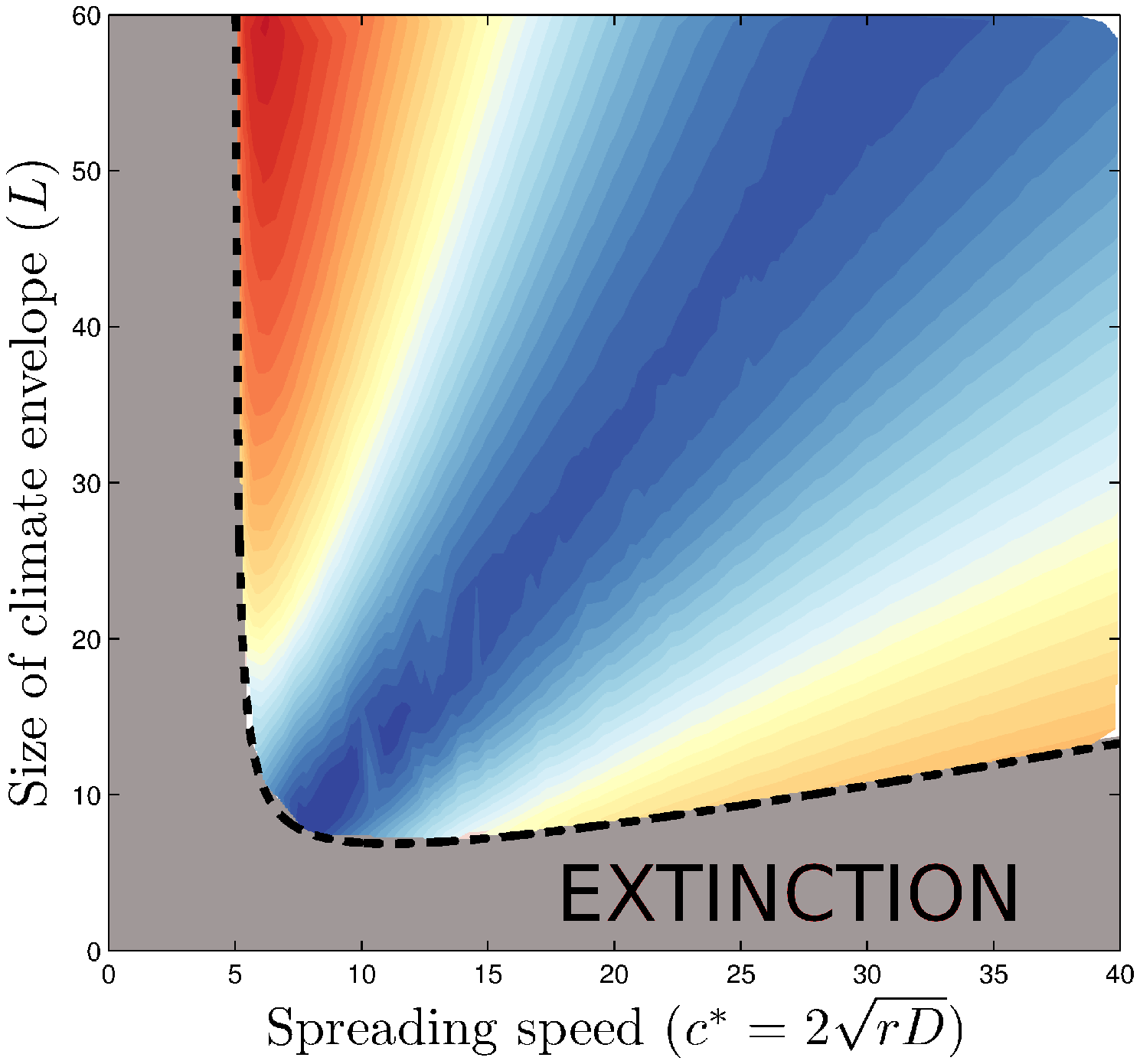}}
\caption{The asymptotic diversity profiles $\Div^2_\infty$ as functions of the spreading speed $c^*:=2\sqrt{rD} \in [c,40]$ and the size of the climate envelope $L\in[L^*(c^*,c),60]$. The profiles are shown with a fixed climate envelope: (a) $c=0$; and for two values of the climate velocity $c$: (a) $c=2$, (b) $c=5$. The white region corresponds to the parameter range where the population faces extinction, and the dashed curve represents the critical size $L^*$ of the climate envelope as a function of the potential spreading speed $c^*$ (see Eq.~\eqref{eq:Lstar_1D} and Fig.~\ref{fig:survival_Lstar}). Diversity reduces as climate velocity increases.}\label{fig:Divq}
\end{figure}

\begin{figure}
\centering
\includegraphics[width=139mm]{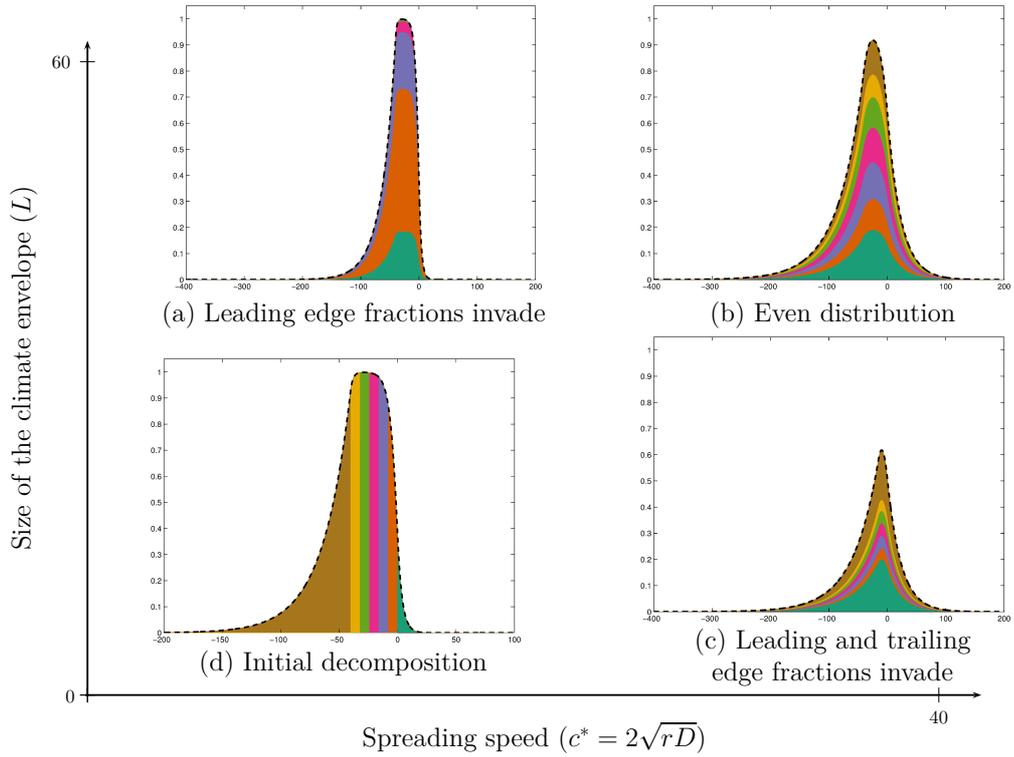}
 \caption{ The spatial genetic patterns inside a range shift (scenario~2). We plot the dynamics of the $7$ fraction densities $\upsilon^i$ inside a population $u,$ described by~\eqref{eq:U_1D} with~\eqref{eq:f_shift}, which follows a climate envelope of finite size $L$, shifting at a slow velocity $c=2.$ We compare the genetic structure of three scenarios from Fig.~\ref{fig:Divq}(b): (a) low spreading speed $c^*=6$ and large envelope $L=40,$ (b) high spreading speed $c^*=20$ and large envelope $L=40$ and (c) high spreading speed $c^*=6$ and small envelope $L=15;$ starting with the initial spatial structure (d). The dashed black curves represent the corresponding travelling wave $U_c.$ }\label{fig:vi_1D_RS}  
\end{figure}

By way of example, we again consider $I=7$ fractions having densities $\upsilon^i$ ($i \in \{1,\ldots,7\}$ inside a population $u$ that follows the climate envelope shifting at speed $c.$ Thus the moving population $u$ is represented by the travelling wave $U_c$. For Figs.~\ref{fig:Divq} and~\ref{fig:vi_1D_RS}, at time $t=0$, $u_0(x)=U_c(x)$ and the fractions satisfy $\upsilon^1_0=U_c\,\mathds{1}_{(-\infty,-L]}$, $\upsilon^i_0=U_c\,\mathds{1}_{(x_{i-1},x_{i}]}$ for $i=2,\dots,I-1$, and $\upsilon^I_0=U_c\,\mathds{1}_{(0,\infty)}$, where the $x_i$ are evenly spaced points with $-L = x_1 < x_2 < \cdots < x_{I-1} = 0$ (see Fig.\ref{fig:vi_1D_RS}(d)).
We numerically solved~\eqref{eq:syst^i} with the climate velocities $c=0,$ $c = 2$ and $c = 5$, the potential spreading speed $c^*=2\sqrt{rD}$, and the size $L$ of the climate envelope ranging in the corresponding survival area (see Fig.~\ref{fig:survival_Lstar}). The growth rate in the climate envelope was set to $r=0.1$.
Then, using the asymptotic abundances $p^i := p(\upsilon_0^i)$ defined in~\eqref{eq:p} Result~1, we computed the asymptotic diversity measures, $\Div^2_\infty$ associated to the travelling wave $U_c$ and its particular initial fraction decomposition $\upsilon_0^i.$ It is defined by
\begin{equation}\label{eq:Divq_infty}
   \ds  \Div^2_\infty=\left(\sum_{i=1}^I{(p(\upsilon^i_0))^2}\right)^{-1}.
\end{equation}

We first notice from Fig.~\ref{fig:Divq} that, for any climate velocity $c$, the diversity index is lower in the regions where the potential spreading speed $c^*$ is either too low or too high compared with the size of the climate envelope.

When the potential spreading speed is low and the size of the climate envelope is large, only the fractions from the leading edge invade the travelling wave (note that orange and green fractions mainly invade in Fig.~\ref{fig:vi_1D_RS}(a)). As already observed in the range expansion scenario~2, fast climate change increases the weight of the leading edge and gene surfing may occur.
However, diversity loss due to fast colonisation at the leading edge can be offset by a small climate envelope which increases extinction lag and allows the fractions at the bulk of the population to contribute to the travelling wave.

When the potential spreading speed is high compared with the climate velocity $c$, the climate change imposes strong constraints on the expansion of the population. Thus, as expected from our previous numerical result (see Fig.~\ref{fig:vi_1D}), genetic diversity is better conserved at the leading edge of the colonisation wave and all the fractions in the bulk of the population significantly contribute to the wave (see Fig.~\ref{fig:Divq} and Fig.~\ref{fig:vi_1D_RS}(b)).  
However, a population with a very high potential spreading speed $c^*$ can face strong erosion in a small climate envelope (see Fig.~\ref{fig:Divq} and Fig.~\ref{fig:vi_1D_RS}(c)). In this case, the population range is much wider than the climate envelope. Thus the fractions outside the climate envelope have more weight than the fraction in the bulk of the population. Since climate change promotes gene mixing as mentioned above, the fractions at the leading edge and the trailing edge will thus mainly contribute to the wave, which tends to give an uneven distribution of fractions and decreases the genetic diversity measure (note that green and brown fractions invade in Fig.~\ref{fig:vi_1D_RS}(c)). 

\section{Discussion}\label{sec:discusion}
Using a deterministic mathematical model, recognised as a robust descriptor of colonisation waves~\citep{BerDie09,PotLew04} as well as a good predictor of allele dynamics inside a population~\citep{RoqGarHamKle12,Nag75,Nag80}, we first showed that range expansion under climate change (scenario~1) modifies the genetic diversity of the colonisation front. Under climate change, all the fractions of the population are preserved in the colonisation wave, even though the distribution of the proportions may change. In the absence of climate change, only the fractions initially ahead of the population, eventually remain in the colonisation front, indicating a strong erosion of diversity due to demographic advantage  of isolated populations at the leading edge of the colonisation front~\citep{HalNel08,RoqGarHamKle12}.
We also showed that the neutral genetic dynamics of a population shifting its range due to climate change (scenario~2) depends crucially on the climate velocity, the size of the climate envelope as well as on its dispersal or spreading ability. 
While an increase in the climate velocity tends to reduce diversity, the intensity of erosion depends crucially on the ratio between the size of the climate envelope and the potential spreading speed of the species. Using diversity indices, we showed that a species whose potential spreading speed is greater than the climate velocity, but not too much greater, can avoid both extinction and erosion of diversity amid climate change.

Our result modifies and extends the commonly held perspective that expansion processes generally erode the neutral genetic diversity along the colonisation front. It reveals that two typical examples of range expansion, biological invasions by alien organisms and the movement of species in response to climate change, should induce different effects on genetic diversity of the population. 
In the context of biological invasions, the alien organism spreads into a newly suitable habitat without any environmental or climate constraints, so it expands its range in a favourable homogeneous habitat. In this case, an erosion of diversity occurs at the leading edge of the colonisation front~\citep[see also][]{HalNel08,RoqGarHamKle12}. However, we show that a colonisation or recolonisation driven by climate change could promote diversity along the travelling wave of colonisation~\citep[see also][]{Plu11,DaiXiaLuFu14,NulHal13}.

This result also reveals the important role played by climate envelopes, arising from climate conditions, in the preservation of genetic diversity. They represent the climate constraints imposed on migration. 
Ahead of the envelope, the growth rate of individuals is negative, reducing the speed of propagation of the species and allowing a diversity of genes from the centre or the trailing edge of the range to reach the leading edge, as shown in Result~\ref{theo:pushed}. This effect of climate envelopes has already been observed in the empirical literature. For instance,~\citet{Plu11} investigated the genetic diversity of a population of European larch whose spread behind a retreating glacier showed a high level of genetic diversity at the leading edge of the range. The reduction of the growth rate ahead of the climate envelope, where the population density is low, is conceptually similar to the sink produced by an Allee effect, which is also known to promote diversity in a travelling wave of colonisation~\citep{RoqGarHamKle12}. In that sense, our results generalise the idea that any mechanism that either provides a population sink ahead of a moving population or constraint the expansion should help the population 
to maintain its initial genetic diversity. 

Behind the envelope, the growth rate is also negative, providing a contraction constraint on the species' range. 
Counter--intuitively, increasing contraction constraint enlarges the species range region located behind the climate envelope. Thus, the fractions from the rear edge of the travelling wave are pushed to survive and invade the leading edge of the wave when the contraction constraint is high (see Fig.~\ref{fig:vi_1D_RS}(c)). 
The beneficial effect of such a contraction constraint on preservation of initial level of diversity has also been investigated by others~\citep{AreRay12,LebEstStr06}. Overall, the mixing process induced by the climate envelope is a defining characteristic of \emph{pushed} travelling waves~\citep{BonCovGarHamRoq14}. In that sense, our results extend the idea that any process that create pushed travelling waves should allow the population to spatially promote its initial genetic diversity.

Genetic drift is not explicitly modelled in our forward approach, and this is an important difference between this study and~\citep{DaiXiaLuFu14,NulHal13}. In their study, the critical role of random genetic drift eventually leads to the fixation of a single  allele in the travelling wave, causing a total loss of genetic diversity. However, consistent results are obtained from their backward stochastic approach. Moreover, recent study~\citep{DurWai16} has shown that the coalescent stochastic model used in~\citep{DaiXiaLuFu14,NulHal13} is well represented by our deterministic model in which random genetic drift disappears because it has been averaged when the local population size is large enough.


Dispersal ability is fundamental in determining genetic diversity responses to climate change. However, recent evidence has shown that climate change can also impact species dispersal during range shifts, either directly or indirectly~\citep{TraDelBoc13}. Theoretical evolutionary studies predict increased  dispersal at expanding range margins, matching observations made for various species~\citep{CwyMcD87,SimTho04,HilThoBla99,TraMusBenDyt09,Bal06,BurPhiTra10,HenBocTra13,PerPhilBasHas13}. However, few mathematical studies have investigated the evolution of species dispersal and its impact on the spatial structure of a species' range when expansions occur across environmental gradients~\citep{PeaLan89,KubHovPoe10,Phi12}. Our analysis suggests that increased dispersal ability during range shifts or range expansion due to changing climate envelopes increases potential spreading speed of the species and may enhance genetic diversity. As already pointed out by pioneering works of~\citet{TraMusBenDyt09} and~\citet{
Phi12}, we will need to integrate ecology and evolution in  order to understand the complex, intertwined effects of dispersal and climate change on genetic diversity.

Our study provides tools that project genetic diversity measure from the climate velocity, the potential spreading speed of the species, and the size of the climate envelope data. The last IPCC report~\citep{IPCC14} predicts a climate velocity of $c=2\,\si{km.yr^{-1}}$ on average under the worst climate change scenario (RCP 8.5). On the other hand, the mean potential spreading speed of several terrestrial and freshwater species range from $c^*=0.1\,\si{km.yr^{-1}}$, for some trees and herbaceous plants to $9\,\si{km.yr^{-1}}$ for artiodactyls. Moreover, their average range size varies widely from $L=1.2\,\si{km}$ for fish and vascular plants to $L=60\,\si{km}$ for birds and mammals~\citep{BroSteKau96}, although there is a high degree of variation around these means. We can thus see from our results in Fig.~\ref{fig:Divq} which species may be in danger of extinction and loss of genetic diversity.

From a mathematical standpoint, our study contributes a new insight on the extensively studied topic of travelling wave solutions of reaction--diffusion equations in heterogeneous environments. Following the recent approach developed by~\citet{GarGilHamRoq12} to characterise the pulled--pushed nature of travelling wave, we focus on the dynamics of the inside structure of travelling waves, conversely to classical approaches analysing the dynamics of the total waves.
Our result generalises techniques developed in~\citep{GarGilHamRoq12} to a large class of equations and shows that travelling wave solutions of reaction--diffusion with a spatial heterogeneous environment developing in time are pushed in the sense of ~\citep{GarGilHamRoq12}. These flexible and intuitive mathematical techniques could be used to more complex models that do not necessarily admit travelling wave solutions. For instance, in a two-dimensional environment, species expanding their range in response to climate change can face absolute boundaries to dispersal because of external environmental factors. Consequently, not only the position but also the shape and size of the climate envelope can change with time. In this case, travelling wave solutions may not exists but the mathematical techniques can still apply.

\section{Proof of Result~\ref{theo:pushed}}\label{sec:proof}
First of all, using the substitution $u_\mathrm{new}(t,x)=u(t/r,x\sqrt{D/r})$, we can assume that the diffusion coefficient $D$ is equal to $1$. Now, in the moving frame having speed $c$, the fraction density $\upsilon$ can be written $\tilde \upsilon(t,x)=\upsilon(t,x+ct)$ and the time dependence of the reaction term vanishes. We can use the Liouville transform $\upsilon^*(t,x)=\tilde\upsilon(t,x)e^{cx/2}$ to remove the advection terms. Thus, the function $\upsilon^*$ satisfies the linear equation
   \begin{equation}\label{eq:vstar}
\ds \partial_t \sups=  \partial_{xx} \sups +\sups\,\big(f(x,U_c(x))/U_c(x)-c^2/4\big)
\end{equation}
(in which advection and time dependence are absent) with the initial condition $\sups(0,x)=\upsilon_0(x)e^{c\,x/2}$.

We show that $\upsilon^*$ can be decomposed as the sum of a stationary function and a function that converges to $0$ exponentially as $t\to\infty$. Note that $\varphi(x)=e^{c \, x /2}\, U_c(x)$ is a positive eigenfunction of the operator that appears in the right-hand side of~\eqref{eq:vstar} and that the associated eigenvalue is $0$. On the one hand, if $f$ satisfies~\eqref{eq:f_shift} then since $U_c$ satisfies~\eqref{eq:U_1D}, one can check that $U_c(x)\sim \exp(-\gamma|x|)$ as $|x|\to\infty$, where $\gamma:=\sqrt{1+c^2/4}$. On the other hand, if $f$ satisfies~\eqref{eq:f_infty} we have $U_c(x)\sim \exp(-\gamma x)$ as $x\to+\infty$ and $U_c(x)\to 1$ as $x\to-\infty.$ In both cases, Sturm--Liouville theory implies that $0$ is the largest eigenvalue of this operator, the remainder of the spectrum being located to the left of some negative constant $-\mu$. Thus, we can write 
\begin{equation}
\upsilon^*(t,x)=p\, \varphi(x) +z(t,x),\vspace{-0.1cm} \label{eq:decompu*}
\end{equation}
where $p \in \R$ and $z$ is orthogonal to $\varphi$ in the sense that $\ds{\int_{-\infty}^{\infty} z(t,x) \varphi(x) \, dx =0}$ for each $t\ge 0$. And $|z(t,x)|\le K_1 \, e^{-\mu\, t}$ for some constant $K_1>0$. Setting $t = 0$ in the expression in~\eqref{eq:decompu*}, multiplying by $\varphi$, and then integrating, we get the expression in~\eqref{eq:p} for $p$.

Finally, we have $|\upsilon^*(t,x)-p \, \varphi(x)|\le K_1\,  e^{-\mu \,t }$, so $|\tup(t,x)-p\, U_c(x)|\le e^{-cx/2}\, K_1e^{-\mu \,t } \, $ for all $x\in\R$. It follows that $\upsilon(t,x+ct)-pU_c(x)\to 0$ uniformly on compacts as $t\to+\infty$ and even uniformly in any interval of the type $[A,+\infty)$ with $A\in\R.$
In addition, if $f$ verifies~\eqref{eq:f_shift}, we get $|z(t,x)|\leq K_2 e^{-\gamma|x|}$ for some constant $K_2>0$. Thus, $|\upsilon^*(t,x)-p \, \varphi(x)|\le \min{\Big(K_1\,  e^{-\mu \,t },K_2 \,E^{-\gamma|x|}\Big)}$, so $|\tup(t,x)-p\, U_c(x)|\le K\, e^{-\mu(1-c/(2\gamma))t }$ for all $x\in\R$. Since $\gamma>c/2$, this shows that the fraction density $\upsilon$ converges to the proportion $p$ of the total population $u(t,x)=U_c(x- c \, t)$ uniformly in $\R$ as $t\to\infty$.

\paragraph{Acknowledgements.}
{\small   MAL gratefully acknowledges a Canada Research Chair, a Killam Research Fellowship, Discovery and Accelerator grants from the Canadian Natural Sciences and Engineering Research Council, and the Natural Science and Engineering Research Council of Canada (grant no. NET GP 434810-12) to the TRIA Network, with contributions from Alberta Agriculture and Forestry, Foothills Research Institute, Manitoba Conservation and Water Stewardship, Natural Resources Canada - Canadian Forest Service, Northwest Territories Environment and Natural Resources, Ontario Ministry of Natural Resources and Forestry, Saskatchewan Ministry of Environment, West Fraser and Weyerhaeuser. JG gratefully acknowledges the NONLOCAL project from the French National Research Agency (ANR-14-CE25-0013). 
}

\end{document}